\documentclass[adp,a4paper,fleqn]{w-art}
\usepackage{amssymb}
\usepackage{amsmath}
\usepackage{times,cite,w-thm}
\usepackage{graphicx}

\theoremstyle{plain}

\theoremstyle{definition}

\begin{document}
\DOIsuffix{theDOIsuffix}
\Volume{XX}
\Month{XX}
\Year{20XX}
\pagespan{1}{}
\Receiveddate{XXXX}
\Reviseddate{XXXX}
\Accepteddate{XXXX}
\Dateposted{XXXX}
\keywords{General Relativity and Gravitation; Geometry, differential
geometry and topology; Gravity in more than four dimensions, Kaluza-Klein
theory, unified field theories.}
\subjclass[pacs]{04.00.00, 02.40.-k, 04.50.+h
\qquad\parbox[t][2.2\baselineskip][t]{100mm}{%
  \raggedright
\vfill}}%
\title[Algebraic Geometry Approach in Gravity Theory]{Algebraic Geometry
Approach in Gravity Theory and New Relations between the Parameters in Type
I Low-Energy String Theory Action in Theories with Extra Dimensions}
\author[B.G. Dimitrov]{Bogdan G. Dimitrov\inst{1,}%
\footnote{Corresponding author\quad E-mail:~\textsf{bogdan@theor.jinr.ru},
          Phone: 7\,-49621\,62\,445,
          Fax: 7\,49621\,65\,-084}}
\address[\inst{1}]{Bogoliubov Laboratory of Theoretical Physics, Joint
Institute for Nuclear Research, 6 Joliot-Curie str., 141 980, Dubna, Russian Federation}
\begin{abstract}
On the base of the distinction between covariant and contravariant metric
tensor components, a new (multivariable) cubic algebraic equation for
reparametrization invariance of the gravitational Lagrangian has been
derived and parametrized with complicated non - elliptic functions,
depending on the (elliptic) Weierstrass function and its derivative. This is
different from standard algebraic geometry, where only two-dimensional cubic
equations are parametrized with elliptic functions and not multivariable
ones.

Physical applications of the approach have been considered in reference to
theories with extra dimensions. The s.c. "length function" l(x) has been
introduced and found as a solution of quasilinear differential equations in
partial derivatives for two different cases of "compactification +
rescaling" and "rescaling + compactification". New physically important
relations (inequalities) between the parameters in the action are
established, which cannot be derived in the case $l=1$ of the standard
gravitational theory, but should be fulfilled also for that case.
\end{abstract}

\maketitle

\DOIsuffix{theDOIsuffix} 
\Volume{XX} \Month{XX} \Year{20XX} 
\pagespan{1}{} 
\Receiveddate{XXXX} \Reviseddate{XXXX} \Accepteddate{XXXX} \Dateposted{XXXX}


\section{Introduction}

Inhomogeneous cosmological models have been intensively studied in the past
in reference to colliding gravitational \bigskip waves [1] or singularity
structure and generalizations of the Bondi - Tolman and Eardley-Liang-Sachs
metrics [2, 3]. In these models the inhomogeneous metric is called the
Szafron-Szekeres metric [4-7]. In [7], after an integration of one of the
components - $G_{1}^{0}$ of the Einstein's equations, a solution in terms of
an elliptic function is obtained. This is important since valuable
cosmological characteristics for observational cosmology such as the
Hubble's constant $H(t)=\frac{\overset{.}{R}(t)}{R(t)}$ and the deceleration
parameter $q=-\frac{\overset{..}{R}(t)R(t)}{\overset{.}{R}^{2}(t)}$ may be
expressed in terms of the Jacobi's theta function and of the Weierstrass
elliptic function respectively [8]. Also in [7], the expression for the
metric in the Szafron-Szekeres approach has been obtained in terms of the
Weierstrass elliptic function after reducing the component $G_{1}^{0}$ of
the Einstein's equations [7, 8] to the nonlinear differential equation $%
\left( \frac{\partial \Phi }{\partial t}\right) ^{2}=-K(z)+2M(z)\Phi ^{-1}+%
\frac{1}{3}\Lambda \Phi ^{2}$. Then by introducing some notations this
equation can be brought to the two-dimensional cubic algebraic equation $%
y^{2}=4x^{3}-g_{2}x-g_{3}$, which according to the standard algebraic
geometry prescription (see [9] for a contemporary introduction into
algebraic geometry)\ can be parametrized as $x=\rho (z)\text{ \ \ },\text{ \
}y=\rho ^{^{\prime }}(z)$, where $\rho (z)$ is the well-known Weierstrass
elliptic function
\begin{equation}
\rho (z)=\frac{1}{z^{2}}+\sum\limits_{\omega }\left[ \frac{1}{(z-\omega )^{2}%
}-\frac{1}{\omega ^{2}}\right]  \tag{1.1}
\end{equation}%
and the summation is over the poles in the complex plane. According to the
standard definition of an elliptic curve [9], such a parametrization is
possible if the functions $g_{2}$ and $g_{3}$ are equal to the s.c.
"Eisenstein series" (definite complex numbers) $g_{2}=60\sum\limits_{\omega
\subset \Gamma }\frac{1}{\omega ^{4}};\text{ \ \ }g_{3}=140\sum\limits_{%
\omega \subset \Gamma }\frac{1}{\omega ^{6}}$.

The main goal of the present paper and of the preceeding ones [10, 11] is to
propose a new algebraic geometry approach for finding new solutions of the
Einstein's equations by representing them in an algebraic form. The approach
is based essentially on the s.c. gravitational theory with covariant and
contravariant metrics and connections (GTCCMC) [12], which makes a clear
distinction between covariant $g_{ij}$ and contravariant metric tensor
components $\widetilde{g}^{is}$. This means that $\widetilde{g}^{is}$ should
not be considered to be the inverse ones to the covariant components $g_{ij}$%
, consequently $\widetilde{g}^{is}g_{im}\equiv f_{m}^{s}(\mathbf{x})$. In
the special case when $f_{m}^{s}(\mathbf{x})=l(x)\delta _{m}^{s}$, important
new relations in the form of inequalities will be found between the
parameters in the type I low energy string theory action - the string
coupling constant $\lambda $, the string scale $m_{s}$ (which in these
theories is identified with $m_{grav.}$) \ and the electromagnetic coupling
constant $g_{4}$.

\section{New Algebraic Geometry Approach in Gravity Theory. Embedded
Sequence of \ Cubic \ Algebraic Equations.}

In the framework of the GTCCMC and the distinction between covariant and
contravariant metric tensor components, we shall assume that the
contravariant metric tensor can be represented in the form of the factorized
product $\widetilde{g}^{ij}=dX^{i}dX^{j}$, where the differentials $dX^{i}$
remain in the tangent space $T_{X}$ of the defined on the initially given
manifold generalized coordinates $X^{i}=X^{i}(x_{1},x_{2},....,x_{n})$. The
existence of different from $g^{ij}$ contravariant metric tensor components $%
\widetilde{g}^{ij}$ means that another connection $\widetilde{\Gamma }%
_{kl}^{s}\equiv \widetilde{g}^{is}\Gamma _{i;kl}=\widetilde{g}%
^{is}g_{im}\Gamma _{kl}^{m}=\frac{1}{2}\widetilde{g}%
^{is}(g_{ik,l}+g_{il,k}-g_{kl,i})$, not consistent with the initial metric $%
g_{ij}$, can be introduced. By substituting $\widetilde{\Gamma }_{kl}^{s}$
in the expression for the "tilda" Ricci tensor $\widetilde{R}_{ij}$ and
requiring the equality of the "tilda" scalar curvature $\widetilde{R}$ with
the usual one $R$, i.e. $\widetilde{R}=R$ (assuming also that $\widetilde{R}%
_{ij}=R_{ij}$), one can obtain the s.c. "cubic algebraic equation for
reparametrization invariance of the gravitational Lagrangian" [10, 11]
\begin{equation}
dX^{i}dX^{l}\left( p\Gamma _{il}^{r}g_{kr}dX^{k}-\Gamma
_{ik}^{r}g_{lr}d^{2}X^{k}-\Gamma _{l(i}^{r}g_{k)r}d^{2}X^{k}\right)
-dX^{i}dX^{l}R_{il}=0\text{ \ \ \ .}  \tag{2.1}
\end{equation}

In the same way, assuming the contravariant metric tensor components to be
equal to the "tilda" ones, the Einstein's equations in vacuum were derived
in the general case for arbitrary $\widetilde{g}^{ij}$, when the assumption $%
\widetilde{g}^{ij}=dX^{i}dX^{j}$ is no longer implemented [11].

Now we shall briefly explain the essence of the s.c. method of \ "embedded
sequence of cubic algebraic equations", proposed for the first time in the
paper [11] and enabling to find solutions of multivariable cubic equations
in terms of (non-elliptic) functions, depending on the Weierstrass function
and its derivative. The method is based on representing (the
three-dimensional case is taken as a model example) the initial cubic
algebraic equation (2.1) as a cubic equation with respect to the variable $%
dX^{3}$ only and applying with respect to it the linear - fractional
transformation. Thus a cubic algebraic equation with respect to the
two-dimensional algebraic variety of the (remaining)\ variables $dX^{1}$ and
$dX^{2}$ is derived (further $\alpha $, $\beta $, $\gamma =1$%
, $2$)  \ $p\Gamma _{\gamma (\alpha }^{r}g_{\beta )r}dX^{\gamma }dX^{\alpha
}dX^{\beta }+K_{\alpha \beta }^{(1)}dX^{\alpha }dX^{\beta }+K_{\alpha
}^{(2)}dX^{\alpha }+2p\left( \frac{a_{3}}{c_{3}}\right) ^{3}\Gamma
_{33}^{r}g_{3r}=0$.  From the last equation the solutions of the initially
given multivariable equation (2.1) (called "the embedding equation of the
preceeding one) are found to be [11] \
\begin{equation}
dX^{3}=\frac{\frac{b_{3}}{c_{3}}+\frac{\rho (z)\rho ^{^{\prime }}(z)}{\sqrt{%
k_{3}}\sqrt{C_{3}}}-L_{1}^{(3)}\frac{B_{3}}{C_{3}}\rho (z)-L_{2}^{(3)}\rho
(z)}{\frac{d_{3}}{c_{3}}+\frac{\rho ^{^{\prime }}(z)}{\sqrt{k_{3}}\sqrt{C_{3}%
}}-L_{1}^{(3)}\frac{B_{3}}{C_{3}}-L_{2}^{(3)}}\text{ \ , \ \ \ }dX^{2}=\frac{%
A}{B}\text{ \ \ \ \ \ ,}  \tag{2.2}
\end{equation}%
\begin{equation}
dX^{1}=\frac{\frac{1}{\sqrt{k_{1}}}\rho (z)\rho ^{^{\prime }}(z)\sqrt{%
F_{1}\rho ^{2}+F_{2}\rho (z)+K_{1}^{(2)}}+f_{1}\rho ^{3}+f_{2}\rho
^{2}+f_{3}\rho +f_{4}}{\frac{1}{\sqrt{k_{1}}}\rho ^{^{\prime }}(z)\sqrt{%
F_{1}\rho ^{2}(z)+F_{2}\rho (z)+K_{1}^{(2)}}+\widetilde{g}_{1}\rho ^{2}(z)+%
\widetilde{g}_{2}\rho (z)+\widetilde{g}_{3}}\text{ \ \ \ \ ,}  \tag{2.3}
\end{equation}
\ where in (2.2)  $A:=$ $\frac{1}{\sqrt{k_{2}}}\rho (z)\rho ^{^{\prime }}(z)%
\sqrt{C_{2}}+h_{1}(dX^{1})^{2}+h_{2}(dX^{1})+h_{3}$ and $B:=\frac{1}{\sqrt{%
k_{2}}}\rho ^{^{\prime }}(z)\sqrt{C_{2}}%
+l_{1}(dX^{1})^{2}+l_{2}(dX^{1})+l_{3}$. The found solutions do not
represent elliptic functions, since they cannot be represented in the form $%
dX^{1}=K_{1}(\rho )+\rho ^{^{\prime }}(z)K_{2}(\rho )$, where $\rho $ is the
Weierstrass elliptic function (1.1). Also, since the solution $dX^{2}$
contains in itself $dX^{1}$, it is called "the embedding solution" of $\
dX^{1}$ [11]. Similarly, $dX^{3}$ is the embedding solution of $dX^{1}$ and $%
dX^{2}$.

\section{"Compactification + Rescaling" \ and "Rescaling +\
Compactification" in Type I String Theory.Tensor Length \ Scale.}

The standard approach in type I string theory in ten dimensions is based on
the low - energy action [14, 15, 16]
\begin{equation}
S=\int d^{10}x\left( \frac{m_{s}^{8}}{(2\pi )^{7}\lambda ^{2}}R+\frac{1}{4}%
\frac{m_{s}^{6}}{(2\pi )^{7}\lambda }F^{2}+...\right) \text{ }=\int
d^{4}xV_{6}L\text{\ \ \ \ ,}  \tag{3.1}
\end{equation}
where $L$ is the expression inside the bracket. After compactification to $4$
dimensions on a manifold of volume $V_{6}$, one can identify the resulting
coefficients in front of the $R$ and $\frac{1}{4}F^{2}$ terms with $%
M_{(4)}^{2}$ and $\frac{1}{g_{4}^{2}}$ and obtain as a result the relations
[14] $M_{(4)}^{2}=\frac{(2\pi )^{7}}{V_{6}m_{s}^{4}g_{4}^{2}}$ \ , \ $%
\lambda =\frac{g_{4}^{2}V_{6}m_{s}^{6}}{(2\pi )^{7}}$. The essence of the
proposed new approach in the paper [13] is that the operation of
compactification is "supplemented" by the additional operation of
"rescaling" of the contravariant metric tensor components in the sense,
clarified in Section 2. This means that\ since the contraction of the
covariant metric tensor $g_{ij}$ with the contravariant one $\widetilde{g}%
^{jk}=dX^{j}dX^{k}$ gives exactly (when $i=k$) the length interval $%
l=ds^{2}=g_{ij}dX^{j}dX^{i}$, then naturally for $i\neq k$ the contraction
will give a (mixed) tensor $l_{i}^{k}=g_{ij}dX^{j}dX^{k}$, which can be
interpreted as a \textit{\textquotedblright tensor\textquotedblright\ length
scale} for the different directions. Further the case of general
contravariant tensor components $\widetilde{g}^{is}$ had been assumed when $%
\widetilde{g}^{is}g_{im}\equiv f_{m}^{s}(\mathbf{x}):=l_{m}^{s}:=l\delta
_{m}^{s}$, from where $\widetilde{g}^{is}$ and the \textquotedblright
rescaled\textquotedblright\ scalar quantities $\widetilde{R}$ and $%
\widetilde{F}^{2}$ can easily be expressed [13]. In the following one can
discern two cases:

1st case - "compactification + rescaling". One starts from the
\textquotedblright unrescaled\textquotedblright\ ten - dimensional action
(3.1), then performs a compactification to a five - dimensional manifold and
afterwards\textbf{\ }a transition to the usual \textquotedblright
rescaled\textquotedblright\ scalar quantities $\widetilde{R}$ and $%
\widetilde{F}^{2}$. Then it is required that the \textquotedblright
unrescaled\textquotedblright\ ten - dimensional effective action (3.1) is
equivalent to the five - dimensional effective action after
compactification, but in terms of the rescaled quantities $\widetilde{R}$
and $\widetilde{F}^{2}$ in the right-hand side (R. H. S.) of (3.1). This can
be expressed as follows \
\begin{equation}
S=\int d^{10}x\left( \frac{m_{s}^{8}}{(2\pi )^{7}\lambda ^{2}}R+\frac{1}{4}%
\frac{m_{s}^{6}}{(2\pi )^{7}\lambda }F^{2}\right) =\int d^{4}x\left(
M_{(4)}^{2}\widetilde{R}+\frac{1}{4}\frac{1}{g_{4}^{2}}\widetilde{F}%
^{2}\right) \text{ \ \ \ \ .}  \tag{3.2}
\end{equation}%
Note also that since $R^{(5)}=R^{(4)}$ ( $R^{(5)}$ means the curvature of
the $5D-$spacetime), the compactification is in fact to four dimensions and
consequently the integration in the R. H. S. of (3.2) is over a $4D-$volume.
Expressing the tilda (rescaled) quantities $\widetilde{R}$ and $\widetilde{F}%
^{2}$ in the right - hand side of \ (3.2) \ through the unrescaled ones $R$
and $F^{2}$ by means of the relation $\ \ \widetilde{g}^{is}=l$ $g^{is}$ and
identifying the expressions in front of the \textquotedblright
unrescaled\textquotedblright\ scalar quantities $F^{2}$ and $R\,\ $in both
sides of (3.2), one obtains an algebraic relation and a quasilinear
differential equation in partial derivatives with respect to the length
function $l(x)$ [13].

2nd case - "rescaling + compactification". This case is just the opposite to
the previous one in the sense that the "rescaled" components become
unrescaled ones and vice versa. In an analogous way, an algebraic relation
can be obtained again after comparing the coefficient functions [13], from
where after introducing the notation $\beta \equiv \left[ \frac{(2\pi )^{7}}{%
V_{6}m_{s}^{4}g_{4}^{4}}-M_{(4)}^{2}\right] m_{s}^{4}V_{6}\frac{2}{(2\pi
)^{7}}$ \ and assuming a small deviation from the relation $M_{(4)}^{2}=%
\frac{(2\pi )^{7}}{V_{6}m_{s}^{4}g_{4}^{2}}$ , i.e. $\beta \ll 1$, one can
express the length scale $l(x)$ as $l^{2}=\frac{1}{1-\beta \frac{R}{%
g^{AC}g^{BD}P}}\approx 1+\beta \frac{R}{g^{AC}g^{BD}P}$. In the last
expression $P$ denotes the term with the second partial derivatives of the
metric tensor, i.e $P:=(g_{AD,BC}+g_{BC,AD}-g_{AC,BD}-g_{BD,AC})$. For $l=1$
(when $\widetilde{g}^{is}=g^{is}$ and $\widetilde{g}^{is}g_{im}\equiv \delta
_{m}^{s}$), as should be expected, we can obtain the usual relation $%
M_{(4)}^{2}=\frac{(2\pi )^{7}}{V_{6}m_{s}^{4}g_{4}^{2}}$ . \textit{The above
result may have also an important physical meaning - any deviations from the
relation }$M_{(4)}^{2}=\frac{(2\pi )^{7}}{V_{6}m_{s}^{4}g_{4}^{2}}$\textit{\
}may be attributed to deviations from the usual scale $l=1$ for the standard
gravity theory.

3rd case - simultaneous fulfillment of \textit{\textquotedblright rescaling
+ compactification\textquotedblright\ and \textquotedblright
compactification + rescaling\textquotedblright }. This means that it does
not matter in what sequence the two operations are performed, i.e. the
process of compactification is accompanied by rescaling. From the
simultaneous fulfillment of the two differential equations one obtains a
cubic algebraic equation with respect to $l^{2}$, from where under the
assumption about the positivity of the square of the length function $l(x)$
(consequently - positive roots of the cubic equation), the following two
inequalities (written for compactness as one - the upper and lower signs in $%
\gtrless $ and $\pm $ mean two separate cases), relating the parameters in
the low energy type I string theory action are obtained
\begin{equation}
p^{2}=\frac{b^{2}}{2}+\frac{a^{3}}{27}-b\sqrt{\frac{b^{2}}{2}+\frac{a^{3}}{27%
}}\gtrless \left[ \frac{a_{1}+6a}{18}\pm \frac{a_{1}}{18}\sqrt{a_{1}^{2}+12a}%
\right] ^{3}\text{ \ ,}  \tag{3.3}
\end{equation}%
where $a\equiv a_{2}-\frac{a_{1}^{2}}{3}$ \ \ , \ $b\equiv 2\frac{a_{1}^{3}}{%
27}-\frac{a_{1}a_{2}}{3}+a_{3}$ \ , $Q\equiv \frac{g^{AC}g_{BD}(2\pi
)^{7}g_{4}^{4}P}{\left[ g^{AC}g^{BD}P(2\pi )^{7}g_{4}^{4}-2R\left(
(2\pi )^{7}-M_{4}^{2}V_{6}m_{s}^{4}g_{4}^{4}\right) \right] }$ , $%
a_{1}\equiv -\frac{(2\pi )^{7}}{M_{4}^{2}m_{s}^{4}V_{6}g_{4}^{2}}$ ,\ $%
a_{2}\equiv \frac{(2\pi )^{7}}{M_{4}^{2}m_{s}^{4}V_{6}g_{4}^{2}Q^{2}}$ \ \
and \ \ $a_{3}\equiv -\frac{g_{4}^{2}}{Q^{2}}$ . Therefore, we do not
consider here the case of an imaginary length function $l(x)$ in the case of
the imaginary Lobachevsky space [17], realized by all the straight lines
outside the absolute cone (on which the scalar product is zero, i.e. $%
[x,x]=0 $). \textit{The last two relations (3.3) are new (although rather
complicated) inequalities between the parameters in the low-energy type I
string theory action, which cannot be obtained in the framework of the usual
gravity theory. Also, it is important to stress that since the scale
function }$l(x)$ does not enter in them, they are valid for the standard
gravity theory approach in theories with extra dimensions.
%
%

\end{document}